# Conservation of Charge and Conservation of Current

Bob Eisenberg

October 19, 2016

Version 3: typos corrected and a section is added *"Why does this matter?"* (see p. 10)

*File: Conservation Of Charge And Conservation Of Current October 18-1 2016.Docx*

Available on arXiv as https://arxiv.org/abs/1609.09175

**Abstract**

Conservation of current and conservation of charge are nearly the same thing: when enough is known about charge movement, conservation of current can be derived from conservation of charge, in ideal dielectrics, for example. Conservation of current is enforced implicitly in ideal dielectrics by theories that conserve charge. But charge movement in real materials like semiconductors or ionic solutions is never ideal; indeed, charge often is moved by forces like diffusion, convection, or even heat flow, and it varies a great deal (by a factor of 40 times) in the time domain of importance in important systems. The flow of charge in semiconductors allow our modern digital technology. The flow of charge in ionic solutions are involved in most of electrochemistry and battery technology. Life occurs in ionic solutions within and outside biological cells. In these systems, conservation of current remains true, even though derivations involve unrealistic ideal dielectric coefficients. We present an apparently universal derivation of conservation of current and advocate using that conservation law explicitly as a distinct part of theories and calculations of charge movement in complex fluids and environments. Classical models using ordinary differential equations rarely satisfy conservation of current, including the chemical kinetic models implementing the law of mass action and Markov models. These models must be amended if they are to conserve current. Strict enforcement of conservation of current is likely to aid numerical analysis by preventing artifactual accumulation of charge.

Conservation of charge and conservation of current are closely related. Indeed, when dielectric properties are ideal, conservation of current is naturally derived from conservation of charge as in the text (Jackson 1999) that has taught electrodynamics to generations of physicists. Any theory of an ideal dielectric that conserves charge will conserve current. But materials do not approximate ideal dielectrics in several electrical systems of great technological and biological importance. In semiconductors, dielectric properties vary by factors of 10 (silicon) over the time scales of importance. Dielectric properties are not at all ideal in ionic solutions like seawater or the solutions within and outside biological cells. Effective dielectric constant varies by factors of 40, in the time range of molecular dynamics simulations used widely to connect atomic detail of biological proteins and macroscopic function of living cells and tissues.

Electrochemical engineering and biology occur in complex fluids in which the electric force field joins with other fields (like convection and diffusion and sometimes heat) to move ions. The description of current flow then involves coupled partial differential equations difficult to write consistently in the mathematical sense, so all results are 'transferrable' in the chemical sense of the word. In transferrable models (that are mathematically consistent) all variables satisfy all field and boundary conditions under all circumstances with one set of unchanging parameters. Mathematical consistency is guaranteed by variational methods that deal with energy and dissipation (Ryham, Liu, and Wang 2006; Ryham 2006; Eisenberg, Hyon, and Liu 2010; Horng et al. 2012; Forster 2013; Wu, Lin, and Liu 2014b, 2014a; Xu, Sheng, and Liu 2014; Wu, Lin, and Liu 2015; Wang, Liu, and Tan 2016). But consistency with experimental data is another thing altogether. Mathematically consistent models can be inconsistent physically if (for example) they contain incomplete representations of the significant physics.

This paper deals with complex systems in which dielectric properties are nothing like ideal. We argue that it is wise to include conservation of current as a separate but (nearly) equal conservation law when dealing with such systems. Numerical analysis and simulation are likely to be easier if artifactual charge accumulations are avoided in their discrete representations of continuous differential equations.

## **Classical Derivation of Conservation of Current**

Charge and field are related by the electrostatic equation of Maxwell

$$\text{div } \mathbf{D} = \rho \qquad (1)$$

where we have used the traditional formulation of Maxwell describing the relation of the displacement field and the charge $\rho$ that is not dielectric or polarization charge. The displacement field is defined as

$$\mathbf{D} = \varepsilon_0 \mathbf{E} + \mathbf{P} \qquad (2)$$

$\mathbf{E}$ is the electric field created by any type of charge, dielectric, free or whatever. $\varepsilon_0$ is the electric field constant, the 'permittivity of free space', i.e. a vacuum, that is really constant everywhere and at all times in all conditions, $\mathbf{P}$ is the polarization field as customarily defined so its divergence is the polarization charge $\rho_p$

$$\mathbf{div\,P} = -\rho_p \qquad (3)$$

For ideal dielectrics the most traditional formulation is

$$\mathbf{D} = \varepsilon_r \varepsilon_0 \mathbf{E} \text{ because } \mathbf{P}_{ideal} = (\varepsilon_r - 1)\varepsilon_0 \mathbf{E} \text{ where } \varepsilon_r \text{ is a positive real number, a constant.} \qquad (4)$$

I introduce my own definition of ideal polarization $\mathbf{P}_{ideal}$ that leaves out the polarization of free space $\varepsilon_0 \mathbf{E}$.

The **P** variables of eq. (3)-(4) allow easy display of 'separation of charge' in bulk dielectrics not so apparent if **P** were specified by $\rho_p$ itself, because the divergence theorem relates surface charges to **div P**. These bulk properties of dielectrics were studied in the early history of electrostatics where the difference between separation of charge and conduction of charge was most confusing and important. Separation of charge is most naturally described **P.** Conduction (i.e., flow) of charge is most naturally described by $\rho_p$, $\partial \rho_p / \partial t$, and $\nabla \rho_p$.

Flux **J** and current depend on the velocity of charge motion **u,** whether the charge is described as a set of point charges (delta functions), a continuous distribution of charge, or even as particles of nonzero size which have charge and mass. Some particles have a definite amount of permanent charge (e.g., sodium ions in water) entirely independent of the strength of the electric field. Other particles have polarization charge as well as permanent charge (polarizable ions perhaps including chloride and even calcium). Still other particles (like water molecules themselves) are joined with their neighbors so strongly that their appropriate description as $\rho_p$ and $\rho$ is not yet agreed upon.

$$\mathbf{J} = \rho \mathbf{u}_{mass} \qquad (5)$$

**J** is the flux of charge $\rho$ where (we reiterate) $\rho$ describes all charges of every type that have mass.

Conservation of charges with mass is written as

$$\frac{\partial \rho}{\partial t} + \mathbf{div}(\rho \mathbf{u}_{mass}) = 0 \quad \text{or} \quad \frac{\partial \rho}{\partial t} + \mathbf{div\,J} = 0 \qquad (6)$$

We note that electric current carried by charges (with mass) is $z\mathbf{eJ}$ where z is the number of charges associated with whatever is flowing, e.g., z can be the 'valence' of an ion, +1 for sodium ion), and **e** is the charge on one proton $1.602 \times 10^{-19}$ cou. We do not bother writing the $z\mathbf{e}$ factor when talking about current later in the paper.

We connect flow and charge, by differentiating eq. (1)

$$\frac{\partial \rho}{\partial t} = \varepsilon_0 \,\mathbf{div}\frac{\partial \mathbf{D}}{\partial t} \qquad (7)$$

and from eq. (2)

$$\frac{\partial \rho}{\partial t} = \varepsilon_0 \frac{\partial}{\partial t}\mathbf{div\,E} + \frac{\partial}{\partial t}\mathbf{div\,P} \qquad (8)$$

and introduce flux from eq.(6)

$$-\mathbf{div\,J} = \varepsilon_0 \frac{\partial}{\partial t}\mathbf{div\,E} + \frac{\partial}{\partial t}\mathbf{div\,P} \qquad (9)$$

**Ideal dielectrics with constant dielectric constants** have ideal polarization:

$$\mathbf{P}_{\text{ideal}} = (\varepsilon_r - 1)\varepsilon_0 \mathbf{E} \quad \text{where } \varepsilon_r \geq 1 \text{ is a positive real number} \qquad (10)$$

$\varepsilon_r$ in this paper is the dimensionless (relative) dielectric constant that is a positive real number independent of time, frequency, and other parameters of experiments. We do not generalize the dielectric coefficient into a complex number (see below). Then,

$$-\mathbf{div\,J} = \varepsilon_0 \frac{\partial}{\partial t}\mathbf{div\,E} + (\varepsilon_r - 1)\varepsilon_0 \frac{\partial}{\partial t}\mathbf{div\,E} = \varepsilon_r \varepsilon_0 \frac{\partial}{\partial t}\mathbf{div\,E} \qquad (11)$$

and we have a conservation law for ideal dielectrics as universal as the Maxwell equations from which it is derived.

## Conservation Law: Ideal Dielectrics

$$\mathbf{div}\left(\mathbf{J} + \varepsilon_r \varepsilon_0 \frac{\partial \mathbf{E}}{\partial t}\right) = 0 \qquad (12)$$

We might now define an ideal current

$$\mathbf{I}_{\text{ideal}} = \mathbf{J} + \varepsilon_r \varepsilon_0 \frac{\partial \mathbf{E}}{\partial t}; \quad \varepsilon_r \text{ is a positive real constant} \qquad (13)$$

since $\mathbf{I}_{\text{ideal}}$ is the conserved quantity (although the units are not quite right because $z\mathbf{e}$ is omitted).

**Ideal dielectric constant.** Treating $\varepsilon_r$ as a real positive constant is a far-reaching idealization. The polarization of real materials almost always varies dramatically with time (or frequency of sinusoidal excitation) in the range of experimental interest and often varies in many other ways.

The dielectric 'constant' studied with sinusoidal signals is often generalized into a complex quantity in the literature of impedance spectroscopy (Macdonald 1992; Barsoukov and Macdonald 2005) and the classical literature of dielectrics (Barthel, Buchner, and Münsterer 1995; Buchner and Barthel 2001; Böttcher et al. 1978; Fröhlich 1958). We do not do that here because there is no compact algebraic generalization of the complex dielectric coefficient in the time domain. Convolution type integrals with memory kernels are needed. We thereby also avoid the confusing classical nomenclature that sometimes involves 'admittance' that is not the reciprocal of 'impedance', frequency dependent resistances that are not the reciprocal of (frequency dependent) conductances, imaginary parts of imaginary parts of complex quantities (that turn out to be real numbers), and so on.

**Vacuum dielectrics.** $\varepsilon_r = 1$ is an important special case describing 'free space', including the vacuum between stars and the vacuum within atoms, that occupies most of the space within atoms, because atomic nuclei are so small.

**Conservation of current in applications.** Theories of flow that use only ordinary differential equations in time (like rate theories of mass action and Markov models) often omit the $\varepsilon_r \varepsilon_0\, \partial \mathbf{E}/\partial t$ term in eq. (13) even though the $\varepsilon_r \varepsilon_0\, \partial \mathbf{E}/\partial t$ term arises from the properties of electrodynamics itself and not from the geometry of a particular system (e.g., not just from 'stray capacitances') or from 'self-energy' arising from polarization charge on nearby dielectric boundaries (Nadler, Hollerbach, and Eisenberg 2003).

Current conservation fails in many theories using only ordinary differential equations in time even in the steady state, even when $\varepsilon_r\, \partial \mathbf{E}/\partial t = 0$. The sequence of reactions $\mathbf{A} \rightleftharpoons \mathbf{B} \rightleftharpoons \mathbf{C}$ often have different current for $\mathbf{A} \rightleftharpoons \mathbf{B}$ and for $\mathbf{B} \rightleftharpoons \mathbf{C}$ as can be shown by direct substitution in the equations that usually define the law of mass action or Markov processes, see (Eisenberg 2014b). Un-conserved currents produce artifactual charges. Tiny artifactual charges are likely to have substantial even dramatic effects if they are not quickly removed by another process. The electric field is strong. See p.1.1 third paragraph of (Feynman, Leighton, and Sands 1963).

**Conservation of Matter.** Note the flux of mass $\mathbf{J}_{mass}$ itself is not conserved by these equations (12)-(13). These equations conserve charge, not mass. The conservation of mass must be enforced by a separate theory and then conjoined to electrodynamics. For example, the Navier-Stokes theory of mass flow might be conjoined to the electrodynamic equations for fluids or solutions with charge (Eisenberg, Hyon, and Liu 2010). The resulting interactions can sometimes be simple, but they are often complex and subtle and difficult to describe consistently, so all variables satisfy all equations and boundary conditions with one set of unchanging parameters. An energetic variational approach helps guarantee consistency when treating complex fluids or systems like ionic solutions or ionic liquids.

**Real materials are not ideal.** Real materials have dielectric properties that are nothing like the ideal dielectric defined in eq. (4).

The nonideal properties of dielectrics can be described in many ways. Comparison with the classical literature is made easier if we introduce a new variable called $\mathbf{P}_{excess}$ to describe excess polarization beyond the ideal polarization of eq. (4) and separate it from the vacuum polarization, of the vacuum dielectric and conduction. This formulation allows us to recognize terms in our equations—familiar from the classical literature of impedance spectroscopy and dielectric materials—while describing many of the complex properties of the actual charge movement found in experiments. Our treatment is certainly not general, however, since we assume isotropic properties, and do not deal explicitly with flows developed conjointly from multiple forces, leaving that to a variational analysis of complex fluids.

The actual polarization measured in experiments is then

$$\mathbf{P}_{actual} = \mathbf{P}_{excess} + \mathbf{P}_{ideal} + \mathbf{P}_{vacuum} = \mathbf{P}_{excess} + \underbrace{(\varepsilon_r - 1)\varepsilon_0 \mathbf{E}}_{\text{Ideal Dielectric}} + \underbrace{\varepsilon_0 \mathbf{E}}_{\text{Vacuum Dielectric}} \tag{14}$$

We can write a general expression for the conservation of flux of charge in real matter from the continuity eq. (9) applied to $\mathbf{P}_{actual}$.

$$-\mathbf{div}\,\mathbf{J} = \frac{\partial}{\partial t}\mathbf{div}\left(\mathbf{P}_{excess} + \underbrace{\mathbf{P}_{ideal}}_{\mathbf{div}(\varepsilon_r-1)\varepsilon_0\mathbf{E}} + \underbrace{\mathbf{P}_{vacuum}}_{\mathbf{div}\,\varepsilon_0\mathbf{E}}\right) \quad (15)$$

$$= \frac{\partial}{\partial t}\left(\mathbf{div}\,\mathbf{P}_{excess}\right) + \frac{\partial}{\partial t}\left(\mathbf{div}(\varepsilon_r-1)\varepsilon_0\mathbf{E}\right) + \frac{\partial}{\partial t}\left(\mathbf{div}\,\varepsilon_0\mathbf{E}\right)$$

or in more conventional terms, we write

## **Conservation Law for Actual Materials**

$$\mathbf{div}\left(\mathbf{J} + \frac{\partial \mathbf{P}_{excess}}{\partial t} + (\varepsilon_r-1)\varepsilon_0\frac{\partial \mathbf{E}}{\partial t} + \varepsilon_0\frac{\partial \mathbf{E}}{\partial t}\right) = 0 \quad (16)$$

This actual conservation law involves the excess polarization $\partial \mathbf{P}_{excess}/\partial t$ that has complex properties different in different materials. The law is valid whenever *(a)* Maxwell's equations are true, *(b)* the vector operators can be defined, and *(c)* time and space differentiation can be interchanged.

We might now try to define the current $\mathbf{I}_{actual}$ that is conserved in actual materials as

$$\mathbf{I}_{actual} = \mathbf{J} + \frac{\partial \mathbf{P}_{excess}}{\partial t} + (\varepsilon_r-1)\varepsilon_0\frac{\partial \mathbf{E}}{\partial t} + \varepsilon_0\frac{\partial \mathbf{E}}{\partial t} \quad (17)$$

This current $\mathbf{I}_{actual}$ is conserved universally under all conditions that Maxwell equations are valid.

The definition of 'current' $\mathbf{I}_{actual}$ in eq. (17) seems too vague to use without a constitutive law for $\mathbf{J}$ and $\partial \mathbf{P}_{excess}/\partial t$. A general description seems impossible because $\mathbf{J}$ and $\partial \mathbf{P}_{excess}/\partial t$ are so different in different materials. The conservation law eq. (16) is not of much use, unless $\mathbf{J}$ and $\partial \mathbf{P}_{excess}/\partial t$ are known in detail.

**Real Dielectrics.** The study of real dielectrics (that do not conduct significant current at very long times in steady applied electric fields) is mostly the study of $\mathbf{P}_{excess}$. The excess polarization is very different in different materials, but is almost never negligible. In pure water, it varies from roughly 80 at long times to roughly 2 at the time scales used in molecular dynamics simulations, although the low frequency behavior has considerable complexity when measured with modern methods (Angulo-Sherman and Mercado-Uribe 2011).

**Real Conductors including Dielectrics.** Most systems of interest in chemical technology—for example batteries, desalination and detoxification systems, or nanodevices—contain ions that flow at low frequencies so $\mathbf{J} \neq 0$ even at vanishing frequencies. These devices are not ideal dielectrics.

Ions flow in these systems as a complex function of time because they are driven by migration in the electric field, diffusion in a concentration field, fluid flow in a pressure field, and sometimes heat flow in a temperature field. These fields occur in devices. Devices hardly ever function without power supplied from power supplies. Devices nearly always involve the flow of charges (ions, holes, or electrons) through a complex fluid, like ionic liquids or electrolyte solutions or semiconductors. Living systems, like electrochemical systems and semiconductors, can often form devices in the engineering sense of the word (Eisenberg 2012, 2003). Devices have inputs at one location, outputs at another, and usually use power supplied at still different locations. Analysis requires models that have spatially inhomogeneous boundary conditions and flows. Devices do not exist in the thermodynamic limit of classical statistical mechanics and molecular dynamics.

**A better conservation law is needed** than eq.(12). We seek another form for the conservation law that is more convincing and universal.

A hint of the other form comes if we abandon the separation of $\mathbf{J}$ and $\partial \mathbf{P}_{\text{excess}}/\partial t$ by introducing a new variable for the flux of charge $\tilde{\mathbf{J}}$.

$\tilde{\mathbf{J}}$ is defined as

$$\tilde{\mathbf{J}} = \mathbf{J} + \frac{\partial \mathbf{P}_{\text{excess}}}{\partial t} + \underbrace{(\varepsilon_r - 1)\varepsilon_0 \frac{\partial \mathbf{E}}{\partial t}}_{\text{Ideal Dielectric}} \quad \text{where } \varepsilon_r \geq 1 \text{ is a positive real number} \qquad (18)$$

Vacuum displacement current' $\varepsilon_0\, \partial \mathbf{E}/\partial t$ is not part of $\tilde{\mathbf{J}}$, even though the term $\varepsilon_0\, \partial \mathbf{E}/\partial t$ helps create the magnetic field. The conservation law is then

## More Useful Conservation Law for Actual Materials

$$\mathbf{div}\left(\tilde{\mathbf{J}} + \varepsilon_0 \frac{\partial \mathbf{E}}{\partial t}\right) = 0 \qquad (19)$$

Eq. (19) includes $\tilde{\mathbf{J}}$ and so $\tilde{\mathbf{J}}$ must be defined if it is to be useful. $\tilde{\mathbf{J}}$ is defined using Maxwell's version of Ampere's law.

$$\mathbf{curl}\left(\mathbf{B}/\mu_0\right) = \underbrace{\tilde{\mathbf{J}} + \varepsilon_0 \frac{\partial \mathbf{E}}{\partial t}}_{\text{'Current'}}. \qquad (20)$$

'Current' is defined as anything that produces **curl B** in an experiment. $\mu_0$ and $\varepsilon_0$ are the magnetic and electrostatic constants that are truly constant under all conditions. They specify the scales of electrodynamics and give the velocity of light $= 1/\sqrt{\mu_0 \varepsilon_0}$.

The definition of $\tilde{\mathbf{J}}$ is

$$\tilde{\mathbf{J}} \triangleq \mathbf{curl}\left(\mathbf{B}/\mu_0\right) - \varepsilon_0 \frac{\partial \mathbf{E}}{\partial t} \qquad (21)$$

$\tilde{\mathbf{J}}$ is defined by eq. (21) whenever $\mathbf{curl}\left(\mathbf{B}/\mu_0\right) - \varepsilon_0\, \partial \mathbf{E}/\partial t$ can be measured.

The conservation law follows easily: **div curl = 0** is an identity true whenever **div** and **curl** can be defined so we have the desired derivation of conservation of current eq. (19).

$$\mathbf{div}\underbrace{\left(\tilde{\mathbf{J}} + \varepsilon_0\, \partial\mathbf{E}/\partial t\right)}_{\text{'Current'}} = 0 \qquad (22)\text{ or }(19)$$

In plain language, anything that creates **curl B** is defined as current. Anything that creates **curl B** is conserved. Current creates **curl B**. Current is conserved.

**Summary**: Eq. (21) provides an operational definition of $\tilde{\mathbf{J}}$ that allows $\tilde{\mathbf{J}}$ to be defined by experiments whenever $\mathbf{curl}\left(\mathbf{B}/\mu_0\right) - \varepsilon_0\, \partial\mathbf{E}/\partial t$ can be measured. Eq. (22) shows that conservation of current can be written without reference to the properties of matter.

# Discussion

This paper derives a universal form of the law of conservation of current eq. (19) that does not involve the properties of matter and so is more useful (and convincing) than the form eq. (12) that describes the polarization of matter with a dielectric constant $\varepsilon_r$ assumed to be a constant real number.

**Universal forms of conservation of current** are needed, in my opinion so the many models found in the literature that do not conserve current will be improved. Numerical procedures might run more reliably if conservation of current is universally enforced. Tiny artifactual accumulations of charge in approximate schemes can have significant sometimes explosive effects.

If the conservation of current is viewed as a poor approximation, that depends on the idealization that dielectric coefficients are constants, few scientists will be motivated to change their models. If conservation of current is viewed as a universal law, scientists will realize it must be enforced in their own models. Examples and extensive discussion of models that do not conserve current have been published (Eisenberg 2014b; Eisenberg 2016b) and the numerical consequences of lack of conservation have been estimated as nontrivial.

**Universal Forms are not always needed.** Idealized dielectrics with dielectric constants independent of time and conditions do not need a separate statement of conservation of current no matter how unrealistic that idealization is. In the idealized case, Maxwell's equations and the continuity equation automatically conserve both charge and current. In that ideal case, conservation of charge implies conservation of current. The idealized models do not describe matter very well, but they they conserve current.

Magnetic systems do not need a separate statement of conservation of current because Ampere's law implies conservation of current no matter how badly the models oversimplify the dielectric constant.

**Universal Forms are needed in Realistic Electrostatic Models.** Realistic electrostatic models are important in many systems of practical importance. They include semiconductors, in which magnetic phenomena are not significant under a wide range of conditions. They include ionic solutions like seawater, the electrolytes of electrochemistry, batteries, ionic liquids, and supercapacitors. They include the ionic solutions derived from seawater inside and outside biological cells. An additional statement of conservation of current is needed (in the form of eq. (19)) for these systems, in my view.

Most of these systems are studied with coarse grained models that link the atomic motions of molecular dynamics to macroscopic properties important in electrochemical technology, biology, and the oceans. Coarse grain models and simulations would benefit from a distinct statement of conservation of current. Numerical approximates can only benefit if conservation of current is universally satisfied in numerical schemes or (not quite the same) enforced by a constraint, separate and (nearly) equal in importance to conservation of charge. See speculation below.

# Why does this matter?

Practical consequences are of interest to readers, as they have kindly told me, so I add this section in the third version of this paper and refer the reader to (Eisenberg 2016a; Eisenberg 2016b; Eisenberg 2014b) for further discussion and estimates of the size of effects.

**Example of a resistor.** It is useful to consider flux of charge and flow of current through an Ohm's law resistor. See detailed discussion on p. 13 of (Eisenberg 2016a) and discusion and equaitons on p. 10, **Resistor** section of (Eisenberg 2016b). If only the flux of charge is considered when applying Kirchoff's flux law to a resistor (i.e., Kirchoff's current law without $\varepsilon_0 \, \partial \mathbf{E}/\partial t$), no charge can accumulate and no electrical force or potential difference can develop. A capacitor added artificially to the circuit accumulates charge and forces and potentials develop. (The capacitor is sometimes called a 'stray' capacitance in the engineering literature, but part of the stray capacitance is an unavoidable vacuum capacitor, an essential feature of Maxwell's equations, see Chapters 3, 5, and 10 of (Maxwell 1891).)

The charge accumulation and electrical force and potential difference appear automatically (without adding an artifical capacitor) if one considers the flow of Maxwell's 'current' when applying Kirchoff's current law (including $\varepsilon_0 \, \partial \mathbf{E}/\partial t$ as in eq. (20)). 'Charge accumulation' occurs automatically in the vacuum (in Maxwell's ether, if you wish, described by $\varepsilon_0$, note the subscript is not $r$; see eq. (14)) and nothing needs to be added to the circuit to ensure that forces and potentials exist.

**Models that use ordinary differential equations rarely conserve current**, until amended. Thus, Markov models (of gating in ion channels, for example, (Zheng and Trudeau 2015; Colquhoun and Hawkes 1981; Sakmann and Neher 1995; Magleby and Weiss 1990)) and chemical kinetic models (of enzymes, for example, the classical description in (Dixon and Webb 1979) followed by every textbook and review article I know about), do not satisfy conservation of current, until amended. The consequences are likely to be numerically severe (see appendix of https://arxiv.org/abs/1502.07251 (Eisenberg 2016b)) because enormous forces are created in microseconds if unbalanced currents flow, forces enough in fact to destroy laboratories in many situations, see third paragraph of the first page 1.1 of (Feynman, Leighton, and Sands 1963).

**Models that do not conserve current are not likely to be useful** under more than one set of conditions. Classical models (of chemical kinetics for example) implemented with ordinary differential equations often remedy these difficulties by adjusting parameters to fit data in one set of conditions (Eisenberg 2014b). Such models cannot be transferred (with one set of parameters) to situations with different concentrations or contents of solutions, or with different electric fields, or different current flow, except in most unusual situations, in my view (Eisenberg 2014a, 2014b).

**Thermal Fluctuations of Electrochemical Potential** produce regions of large concentration in which apparently well stirred chemical reactions are likely to occur with properties very different from reactions with average macroscopic concentrations. Thermal fluctuatuions in kinetic energy and electrochemical ptoential produce unavoidable stochastic fluctuations of current (in space and time) that are likely to propagate to macroscopic time scales even in systems with*out* macroscopic flow, as reported by Ferry for example (Ferry 1980).

Stochastic circulating currents need to be included in treatments of chemical reactions at chemical equilibriium, because most of those reactions are likely to occur in small regions with thermal fluctuations in concentrations (i.e., electrochemical potentials) very different from the spatial mean concentrations. Chemical reactions (involving rearrangements of electrons in covalent bonds) are very fast (say $10^{-19}$ sec) and are steep functions of concentration (in most cases). Reactions in such systems will occur in small regions of inhomogeneity (that will often pesist say $10^{-9}$ sec, a very long time compared to the $10^{-18}$ sec needed to complete the rearrangement of electons).

These transient regions of large concentration do not exist in traditional well stirred approximations (Moore and Pearson 1981) Spatial variables do not occur in well stirred approximations, and field theories are approximated by ordinary differential equations in time. The Belousov Zhabotinsky reaction (search the arXiv, Google and Google Scholar for glimpses of the large literature) is the classical example of a phenomenon that cannot be described by the ordinary differential equations (in time) of well stirred approximation. A graphic example is shown at https://www.youtube.com/watch?v=3JAqrRnKFHo. Similar phenomena seem likely to occur more often than not on the nanoscale because of unavoidable thermal fluctuations in electrochemical potential and resulting thermal current flows.

**Speculation concerning numerical methods.** Computations of electrodynamics including diffusion in the general spirit of the Poisson Nernst Planck equations (called drift diffusion in the semiconductor literature) are notoriously difficult when flows and electric fields are large. Electrical fields are large in and near the highly concentrated charges (often > 10 Molar; solid $Na^+Cl^-$ is 37 M) so important in electrochemical applications (near electrodes) and biological systems (in and near ion channels, binding proteins, nucleic acids and enzyme active sites).

It seems likely to me (but admittedly not to others), that some of these difficulties would disappear if current as defined in eq. (22) were forced to be conserved 'exactly' at every step and stage of a numerical approximation and analysis. It seems to me that many of the 'tricks of the trade' (see our enumeration in (Liu and Eisenberg 2015), but of course this is just one listing out of many in the trade) may not be needed if current as defined in eq. (22) were exactly conserved at every step and stage of a numerical analysis, assuming such is possible.